\documentclass[10pt]{article}
\usepackage{fullpage}
\usepackage{amsmath}
\usepackage{graphicx}

\newcommand{\calA}{{\cal A}}
\newcommand{\calB}{{\cal B}}
\newcommand{\calD}{{\cal D}}
\newcommand{\calR}{{\cal R}}
\newcommand{\calL}{{\cal L}}
\newcommand{\calP}{\mathcal{P}}
\newcommand{\calN}{\mathcal{N}}
\newcommand{\calF}{\mathcal{F}}
\newcommand{\calFt}{\tilde{\mathcal{F}}}

\begin{document}

\title{Boundary Conformal Field Theory and Entanglement Entropy in
Two-Dimensional Quantum Lifshitz Critical Point}

\author{Masaki Oshikawa \\
\it Institute for Solid State Physics, University of Tokyo \\
\it 5-1-5 Kashiwanoha, Kashiwa 277-8581 Japan}

\date{July 21, 2010}

\maketitle

\begin{abstract}
I discuss the von Neumann entanglement entropy in
two-dimensional quantum Lifshitz criical point, namely in 
Rokhsar-Kivelson type critical wavefunctions.
I follow the approach proposed by
B. Hsu {\it et al.} [Phys. Rev. B {\bf 79}, 115421 (2009)],
but point out a subtle problem concerning compactification
of replica boson fields:
although one can define a set of new boson fields by
linear combinations of the original fields,
the new fields are not compactified independently.
In order to systematically study boundary conformal field
theory of multicomponent free bosons,
I employ a geometric formulation based on compactification
lattices.
The result from the boundary conformal field theory
agrees exactly with alternative
calculations by J.-M. St\'{e}phan {\it et al.}
[Phys. Rev. B {\bf 80}, 184421 (2009)],
confirming its universality as argued originally
by B. Hsu {\it et al}.
\end{abstract}

\section{Introduction}

Recently, stimulated by developments in quantum information
theory, characterization of quantum many-body systems
in terms of quantum entanglement is studied vigorously.
Entanglement entropy is a quantitative measure of the
quantum entanglement.
A natural problem then, is to divide the system into
two regions and study the entanglement
entropy between them.
In general, short-range correlations, which are non-universal,
give a contribution proportional to the area of the boundary
between the two regions to the entanglement entropy.
This is often called as the area-law contribution.
In one dimension, the boundary consists just of points,
and the area-law contribution is constant with respect
to the system size or to the size of the regions.

Calabrese and Cardy~\cite{CalabreseCardy}
studied the entanglement entropy
between a region of length $L$ and the rest of the system,
in an infinitely long one-dimensional system.
They demonstrated that, in a critical one-dimensional
system described by a conformal field theory (CFT) with
central charge $c$, the entanglement entropy
has a universal contribution $(c/3) \log{L}$.
In addition to being interesting in itself,
this has proved practically useful in determining
the central charge of the system from numerical simulations.

In two (and higher) dimensions, the non-universal area law contribution
diverges and is the leading contribution to the entanglement
entropy, as the size of the region increases.
Nevertheless, there can be universal contributions
as subleading terms in the entanglement entropy.
Kitaev and Preskill~\cite{KitaevPreskill}, and
Levin and Wen~\cite{LevinWen} found that,
in two-dimensional topologically ordered phases,
there is a universal constant term reflecting
the topological order,
in addition to the non-universal area law contribution.
By considering a set of geometries, one can cancel out
the area law contribution to obtain the universal constant.
This is also confirmed numerically.

Classification of critical points in two-dimensional
quantum systems is not as well understood as in
one dimension. Nevertheless, there is a class of
two-dimensional quantum critical points,
for which the universal critical phenomena
can be described precisely.

Some quantum systems in two dimensions can be
related to two-dimensional classical statistical systems,
by identifying the wavefunction in the quantum system
with the statistical probability of the corresponding
configuration in the classical system.
This is a generalization of the Rokhsar-Kivelson
wavefunction introduced for quantum dimer models.~\cite{RokhsarKivelson}

When the classical system is at a critical point,
the quantum system is also at a quantum critical point.
A critical point of two-dimensional classical statistical
systems is often described by a two-dimensional CFT.
If this is the case,
the corresponding quantum critical point
of the quantum system is also governed by the
same CFT.
Let us call such quantum critical points
as two-dimensional conformal critical points,
following Ref.~\cite{FradkinMoore}.
Although such a wavefunction is rather special,
I can expect that a conformal critical point
represents a certain universality class.
An important subclass of such two-dimensional
conformal criticalities is those correspond
to the free boson CFT.
Following Ref.~\cite{Hsu-EE}, I call this
as quantum Lifshitz universality class.
Quantum Lifshitz universality class is
in fact a one-parameter family of universality
class, since the free boson field theory
is characterized by a free parameter
(compactification radius).

Fradkin and Moore~\cite{FradkinMoore}, and
subsequently Hsu, Mulligan, Fradkin, and Kim~\cite{Hsu-EE}
studied the entanglement entropy in
two-dimensional conformal quantum critical points
using replica trick and boundary CFT.
In Ref.~\cite{FradkinMoore}, it was argued that
the entanglement (von Neumann) entropy has the form
\begin{equation}
 S_E = \alpha l - \frac{c}{6} (\Delta \chi)
\ln{\left( \frac{l}{a} \right)} + O(1),
\label{eq:SEwithlog}
\end{equation}
when the boundary $\Gamma$ is smooth.
Here $c$ is the central charge of the CFT,
$\Delta \chi$ is the change in the Euler characteristics
by the partition of the system, $l$ is the length of
the boundary $\Gamma$ between the regions A and B,
$\alpha$ is the non-universal coefficient of the area law
contribution, and $a$ is the ultraviolet cutoff.

When $\Gamma$ is smooth and $\Delta \chi=0$, the logarithmic
term vanishes.
In Ref.~\cite{Hsu-EE}, it was argued that
in such a circumstance, the $O(1)$
term in eq.~\eqref{eq:SEwithlog} contains a universal constant,
similarly to the case of topologically ordered phases
discussed in Refs.~\cite{KitaevPreskill,LevinWen}.

Following these developments,
St\'{e}phan, Furukawa, Misguich, and Pasquier~\cite{Stephan-EE}
studied the same constant term in
eq.~\eqref{eq:SEwithlog} using different analytical
and numerical methods.
They agreed that the constant is a universal quantity
determined by the underlying CFT.
In particular, the universality is confirmed for lattice
models with different microscopic parameters.
However, the universal constant obtained
in Ref.~\cite{Stephan-EE} for the quantum Lifshitz universality
class disagreed with the original prediction
in Ref.~\cite{Hsu-EE}.
The disagreement is potentially serious, since
if both derivations stand valid, it would imply
a breakdown of the universality of the constant term
in the entanglement entropy.

In this paper, I aim to resolve the issue by
re-examining the derivation of the entanglement
entropy in Refs.~\cite{FradkinMoore,Hsu-EE}.
I point out there are subtle problems in
``changing the basis'' technique employed
to derive a fundamental formula
in Refs.~\cite{FradkinMoore,Hsu-EE}.
(I note that, although the fundamental formula
in Refs.~\cite{FradkinMoore,Hsu-EE} does not hold
as an exact identity and the universal constant
reported in Ref.~\cite{Hsu-EE} should be corrected,
the logarithmic term obtained in Ref.~\cite{FradkinMoore}
could still stand valid. I will briefly discuss this point
in Sec.~\ref{sec:conclusion}.)

For a free boson CFT,
new fields defined as linear combinations
of the original field are apparently independent of
each other.
However, they are not completely independent
since the compactification of the new fields
intertwine different components.
This complication is often ignored in literature
and still correct results are obtained in some
cases. However, its negligence can lead to erroneous
results; the present problem of entanglement entropy
is indeed such an example.\footnote{
The problem in the treatment of compactification was pointed out
earlier by V. Pasquier {\it (private communications)},
as a possible source of error in Ref.~\cite{Hsu-EE}. }
The intertwining of new free boson fields in
context of boundary CFT
was discussed by Wong and Affleck~\cite{WongAffleck}
for a quantum impurity problem.
There, the compactification of each component
of new fields is written explicitly in
terms of gluing conditions.
Although taking all the gluing conditions into account
should lead to a correct result, it becomes
increasingly cumbersome for larger number of components.

Instead, the compactification of
multicomponent bosons can be formulated geometrically using
compactification lattices in multidimensional space.
This is useful in construction of some of the possible
boundary conditions systematically, and has been applied
to string theory and to condensed matter physics.
This approach is particularly suitable for the present
problem, as the replica trick requires calculation
for arbitrary number of components.

I demonstrate that the entanglement entropy in
the critical groundstate wavefunction of quantum Lifshitz
univesality class can be indeed calculated fully taking
the above subtlety into account, with the geometrical construction
of the boundary conditions of multicomponent free boson CFT.
As a result, I find that the universal constant
term in the entanglement entropy is in exact
agreement with that obtained in Ref.~\cite{Stephan-EE}
by different approaches.
This agreement confirms the universality of
the constant term, as put forward in Ref.~\cite{Hsu-EE}.

This paper is organized as follows.
In Sec.~\ref{sec:setup}, I quickly review the
arguments in Refs.~\cite{FradkinMoore,Hsu-EE}.
I then point out subtle problems in the
``changing the basis'' trick used in these papers.
In order to demonstrate the problem in the
``changing the basis'' trick, in Sec.~\ref{sec:simple},
I discuss a simple problem of a single component
free boson field theory.
By a ``folding'' trick, the problem can be regarded
as two-component free boson with boundaries.
However, naive application of the ``changing the basis''
technique fails to reproduce the partition function.
In Sec.~\ref{sec:cylinder}, I calculate the entanglement
entropy for the cylinder geometry,
using a geometric formulation of the boson compactification.
The constant term in the entanglement entropy is given
in terms of the universal boundary entropy in
boundary CFT.
The calculation is extended to the torus geometry in
Sec.~\ref{sec:torus}, where it is pointed out that
the entanglement entropy in the torus geometry
is twice of that in the cylinder geometry
for conformal critical points.
Sec.~\ref{sec:conclusion} is devoted to conclusion
and discussions, including whether there would be
any correction the logarithmic
term predicted in Ref.~\cite{FradkinMoore} or not.
The recent e-print by Hsu and Fradkin~\cite{HsuFradkin}
will be also discussed in this section.

Relevant materials in
boundary CFT of multicomponent free
boson field theory are summarized in
Appendix~\ref{app:multiboson}.
In Appendix~\ref{app:simple}, the solution to the
simple example introduced in Sec.~\ref{sec:simple}
is given by explicitly solving the gluing conditions.

\section{Setup of the problem}
\label{sec:setup}

In a class of two-dimensional critical systems, the
groundstate
wavefunction is related to a two-dimensional CFT.
Namely, the ground state is given as
\begin{equation}
 | \Psi_0 \rangle \propto
\int \calD \phi e^{-S[\{ \phi \}]/2} | \{ \phi \} \rangle,
\label{eq:wfCFT}
\end{equation}
for the action $S[\phi]$ of a CFT.

I consider a groundstate given as eq.~\eqref{eq:wfCFT}.
I divide the system into two regions A and B.
The entanglement entropy between the two regions
is defined by the von Neumann
entropy of the subsystem A as
\begin{equation}
 S_E = - \mathrm{Tr} \left( \rho_A \log{\rho_A} \right) ,
\end{equation}
where
\begin{equation}
 \rho_A = \mathrm{Tr}_B | \Psi_0 \rangle \langle \Psi_0 |
\end{equation}

I will follow the argument by Fradkin and Moore to derive
$S_E$ for such a system.~\cite{FradkinMoore}
Namely, the entanglement entropy is rewritten as
\begin{equation}
 S_E = - \frac{\partial \mathrm{Tr}{\rho_A}^n}{\partial n} \big|_{n=1} .
\end{equation}
In the replica trick, we compute $\mathrm{Tr}{\rho_A}^n$ for
an integer $n$ and then make an analytic continuation
to arbitrary integer $n$.
For an integer $n$, we can introduce $n$ copies of the CFT
with the fields $\phi_1, \phi_2, \ldots \phi_n$.
\begin{equation}
 \mathrm{Tr}{\rho_A}^n = \frac{Z_\calP}{Z_F}
\label{eq:tr_rhoAn}
\end{equation} 
where $Z_\calP$ is the partition function 
of the $n$-component field theory with the
condition
\begin{equation}
\phi_1=\phi_2=\ldots = \phi_n
\label{eq:replica-bc}
\end{equation}
at the boundary between A and B.
$Z_F$ is the partition function of the same
$n$-component field theory but without any
restriction at the boundary between A and B.
Both $Z_P$ and $Z_F$ are functions of $n$, although
the dependence is omitted for brevity of the expressions.

Fields with different replica indices are independent,
except possibly at the boundary $\Gamma$ between A and B.
Since no coupling is introduced at the boundary
in $Z_F$, we find
\begin{equation}
 Z_F = \left( z_F \right)^n,
\label{eq:ZnF}
\end{equation}
where $z_F$ is the partition function
of the {\em single} component free boson field theory without
any restriction at the boundary.
I find no problem in the argument, up to this point.

They proceed further by changing the basis,
taking the linear combinations of the original fields
$\phi_j$, as
\begin{align}
 \varphi_0 & = \frac{1}{\sqrt{n}} \sum_{j=1}^n \phi_j 
\label{eq:varphi0}
\\
 \varphi_1 & = \frac{1}{\sqrt{2}} \left( \phi_1 - \phi_2 \right) 
\label{eq:varphi1}
\\
\ldots
\notag \\
 \varphi_{n-1} &= \ldots \notag .
\end{align}
It was argued that, the condition~\eqref{eq:replica-bc}
does not affect the ``center of mass'' field $\varphi_0$
and it thus remains free at $\Gamma$.
On the other hand, 
all the other linear combinations $\varphi_j$ with $j>0$,
which correspond to differences among $\phi_k$,
obey fixed (Dirichlet) boundary condition at $\Gamma$.
The $n-1$ ``difference'' fields was then regarded as independent.
As a consequence of this argument,
it was proposed in Ref.~\cite{FradkinMoore} that
\begin{equation}
 Z_\calP = \left( z_D \right)^{n-1} z_F,
\label{eq:ZnG}
\end{equation}
where $z_D$ is the partition function
for the {\em single} component
field, with Dirichlet boundary condition at the boundary.

Combining eqs.~\eqref{eq:tr_rhoAn},\eqref{eq:ZnF}, and
\eqref{eq:ZnG} leads to
\begin{equation}
\mathrm{Tr}  {\rho_A}^n = \left( \frac{z_D}{z_F}  \right)^{n-1}.
\end{equation}
This implies
\begin{equation}
 S_E = - \log{\frac{z_D}{z_F}} = - \log{\frac{z^A_D z^B_D}{z_F}},
\label{eq:SEinFM}
\end{equation}
where $z^{A}_D (z^B_D)$ is the partition function for the
{\em single} component free boson field theory
restricted to the region A (or B),
with Dirichlet boundary condition at the boundary between A and B.
Eqs.~\eqref{eq:ZnG} and~\eqref{eq:SEinFM} are also the
basis of the calculations in Ref.~\cite{Hsu-EE}.

However, in a general CFT,
the field $\phi$ is subject to interactions.
For example, the two-dimensional critical Ising model
correspond to $\phi^4$ field theory with a certain
fine-tuning.
The field theory with the interaction
is generally not invariant under orthogonal
transformations of the fields.
For example, let us consider the simplest case $n=2$.
The Lagrangian of two identical Ising field theory would read
\begin{equation}
\calL_{{\rm (Ising)}^2} = \sum_{j=1,2} \left(
\frac{1}{2} (\partial_\mu \phi_j)^2 
+ \frac{1}{2} m^2 {\phi_j}^2
+ \frac{\lambda}{4} {\phi_j}^4 \right),
\end{equation}
where $m^2$ and $\lambda$ are fine-tuned to make the
system critical.
Now I introduce the new fields $\varphi_{0,1}$
as in eqs.~\eqref{eq:varphi0} (for $n=2$)
and \eqref{eq:varphi1}.
Then, in terms of the new fields,
\begin{equation}
\calL_{{\rm (Ising)}^2} = \sum_{j=1,2}
\left(
\frac{1}{2} (\partial_\mu \varphi_j)^2 
+ \frac{1}{2} m^2 {\varphi_j}^2  \right)
+ \frac{\lambda}{4} 
\frac{1}{2}
\left( {\varphi_0}^4 +
6 {\varphi_0}^2 {\varphi_1}^2
+ {\varphi_1}^4
\right).
\end{equation}
The new fields $\varphi_{0,1}$ are subject to
different interaction from the original one.
Moreover, different components $\varphi_0$ and $\varphi_1$
are now coupled through the bulk interaction
${\varphi_0}^2 {\varphi_1}^2$ and not independent of
each other.
Thus eq.~\eqref{eq:ZnG} would not hold in general
field theories with interactions.

For a free field theory, on the other hand,
the change of the basis appears more legitimate,
because the theory is also free in terms of
the new basis fields.
Let us now consider the case of the free boson field
theory. 
As I have discussed, this will be relevant for
entanglement entropy in the quantum Lifshitz universality
class.
I define the Lagrangian density following the convention
in Ref.~\cite{YJunction-full} as
\begin{equation}
 \calL = \frac{g}{4\pi} (\partial_\mu \phi)^2 .
\label{eq:Lag}
\end{equation}
The field $\phi$ is subject to compactification, namely
the identification 
\begin{equation}
  \phi \sim \phi + 2 \pi R,
\label{eq:compact}
\end{equation}
where $R$ is the compactification radius.

I note that, there is unfortunately a large variety
of conventions for the free boson field theory,
based on different normalizations.
In fact, by a renormalization of the field $\phi$,
we can fix the value of either coupling constant $g$
or the compactification radius $R$.
However, we cannot fix both $g$ and $R$ by the renormalization.
This leaves one free parameter, which governs the critical
behavior.
We can choose either to fix $g$ and consider $R$ as a free
parameter, or to fix $R$ and regard $g$ as a free parameter.
Both conventions, with various choices of the fixed value,
appear in literature.
In this paper, I keep both $g$ and $R$ as parameters.
This is a redundant parametrization, but makes it easier
to compare with literature by setting either $g$ or $R$
to the convention of each paper.
For example, the convention in Refs.~\cite{Hsu-EE,Stephan-EE}
can be recovered by setting $g=1/2$.

Employing the replica trick, I consider $n$ component
free boson theory, in which each component is independent
in the bulk.
For the free field case, the Lagrangian density 
can be also written as
\begin{equation}
 \calL = \frac{g}{4\pi} \sum_{j=0}^{n-1} (\partial_\mu \varphi_j)^2 .
\end{equation}
That is, the new fields are governed by the same Lagrangian
density as the original fields, and there is no interaction
which couples different components.
Thus the arguments in Refs.~\cite{FradkinMoore,Hsu-EE} appears
to valid.
However, even for the free boson field theory, eq.~\eqref{eq:ZnG}
does not quite hold
because of subtlety in boson compactification.

\section{A simple example}
\label{sec:simple}

In order to illustrate the issue, let me discuss a
simple example.
I consider a single component free boson theory~\eqref{eq:Lag}
on a rectangle of the size $2L \times \beta$, with the periodic
boundary condition on both directions.
In other words, the system is defined on a torus.
The partition function is given by
\begin{align}
 Z_{\rm simple}  &= \frac{1}{\eta{(q)}} \frac{1}{\eta{(\bar{q})}}
 \sum_{n,m = - \infty, \infty}
q^{\frac{1}{4}(\frac{m}{\sqrt{g}R} + n \sqrt{g} R)^2}
\bar{q}^{\frac{1}{4}(\frac{m}{\sqrt{g}R} - n \sqrt{g} R)^2}
\notag \\
 &=
\left(\frac{1}{\eta{(q)}} \right)^2
q^{\frac{1}{2}(\frac{m^2}{gR^2}+n^2 g R^2)} , 
\label{eq:Zsimple}
\end{align}
where
\begin{equation}
 q = e^{- \pi \beta/L} .
\end{equation}
For a general torus, $q$ is given as $q=e^{2\pi i \tau}$
where $\tau$ is the modulus of the torus which is a complex
number, and its complex conjugate $\bar{q}$
is distinguished from $q$.
However, here I only consider the rectangular case
where $\tau$ is a pure imaginary, and hence $q = \bar{q}$.
I have used this fact in the second line of
eq.~\eqref{eq:Zsimple}.

\begin{figure}
\begin{center}
\includegraphics[width=0.6\textwidth]{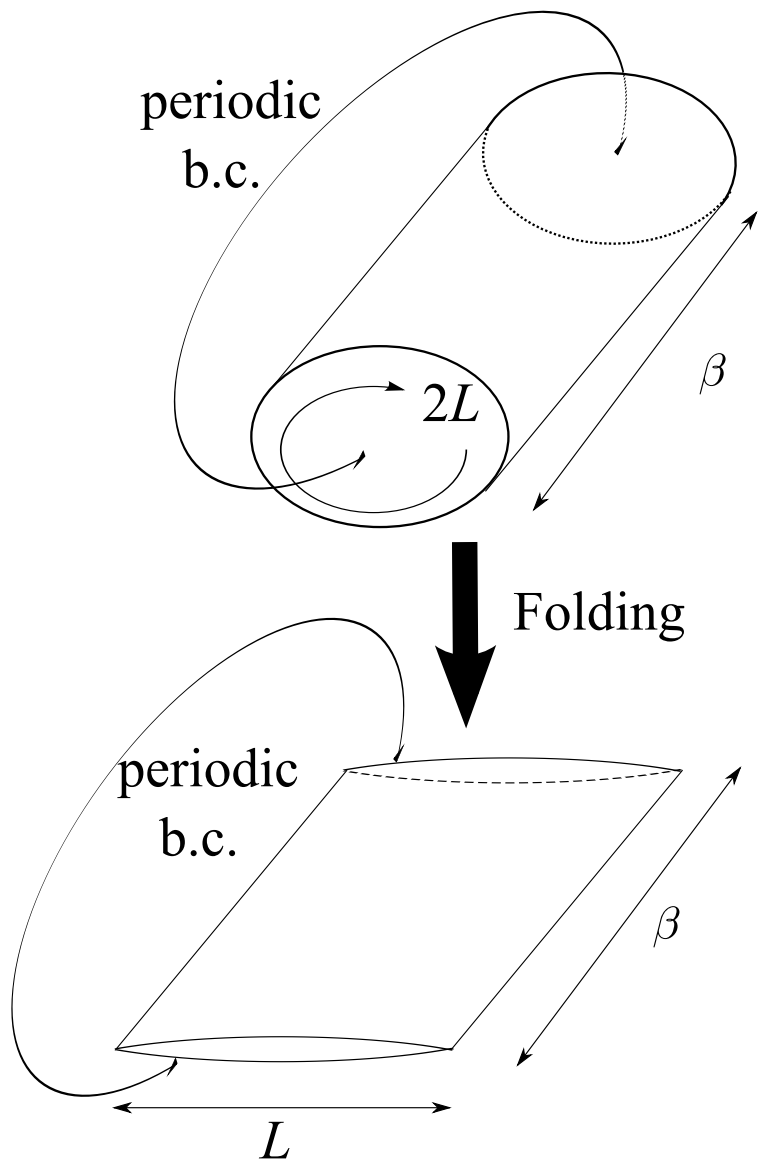}
\end{center}
\caption{A single component free boson field theory is defined on the
torus of size $2L \times \beta$. The system is folded to
a two-component free boson field theory defined on 
the cylinder with circumference $\beta$ and length $L$.
}
\label{fig:simple-fold}
\end{figure}

Now I ``fold'' the system, as shown in Fig.~\ref{fig:simple-fold}.
After the folding, the system can be
regarded as two-component boson field $\phi_1, \phi_2$
defined on a rectangle of size $L \times \beta$.
While the periodic boundary condition is still applied on the
$\beta$ direction, there are two boundaries which were
the folding lines.
Thus the system after the folding is topologically a cylinder.
At the two boundaries, I impose the condition
\begin{equation}
\phi_1 = \phi_2 . 
\label{eq:simple_bc1}
\end{equation}
Namely, if we consider the new fields
\begin{align}
\Phi_0 &= \frac{\phi_1 + \phi_2}{\sqrt{2}}, \\
\Phi_1 &= \frac{\phi_1 - \phi_2}{\sqrt{2}} ,
\label{eq:Phi_simple}
 \end{align}
$\Phi_1$ obeys the Dirichlet boundary condition
$\Phi_1=0$, while $\Phi_0$ remains free (obeys
the Neumann boundary condition) at the two boundaries.

If we apply the same argument as in Ref.~\cite{Hsu-EE},
we would obtain
\begin{equation}
 Z_{\rm simple}(q) = z_{DD}(R,q) z_{NN}(R,q),
\label{eq:wrongsimple}
\end{equation}
where 
\begin{equation}
 z_{DD}(R,q) = \frac{1}{\eta{(q)}} \sum_{n} q^{gn^2 R^2}
 = \frac{1}{\sqrt{2gR}} \frac{1}{\eta{(\tilde{q})}}
  \sum_{n} \tilde{q}^{n^2/(4gR^2)},
\label{eq:ZDD1}
\end{equation}
is the Dirichlet-Dirichlet amplitude for the single component
boson with the compactification radius $R$,
and
\begin{equation}
 z_{NN}(R,q) = \frac{1}{\eta{(q)}} \sum_{n} q^{n^2/(g R^2)}
 = \sqrt{\frac{gR}{2}} \frac{1}{\eta{(\tilde{q})}}
  \sum_{n} \tilde{q}^{g n^2R^2/4},
\label{eq:ZNN1}
\end{equation}
is the Neumann-Neumann amplitude for the same theory.
Here, $\eta{(q)}$
the Dedekind eta function defined in eq.~\eqref{eq:Dedekind_eta}.
However, from eqs.~\eqref{eq:ZDD1} and \eqref{eq:ZNN1}
we can immediately see that eq.~\eqref{eq:wrongsimple}
actually does not hold.

One of the problems is that, in the ``closed string channel'',
the right-hand side of eq.~\eqref{eq:wrongsimple} reads
\begin{equation}
z_{DD}(R,q) z_{NN}(R,q) =
\frac{1}{2} \left(\frac{1}{\eta{(\tilde{q})}}\right)^2
 \sum_{n,m} \tilde{q}^{n^2/(4gR^2)+ m^2 gR^2/4} .
\end{equation}
This implies that each boundary has the
``groundstate degeneracy'' (exponential of the
boundary entropy) $1/\sqrt{2}$, for any value of $R$.
This is in contradiction to the fact that the boundary
in the present example is just a result of an artificial
``folding'' along a line in the bulk,
and thus should not have any boundary entropy.
In fact, the modular invariance of the partition function
on the original torus of $2L \times \beta$ implies
\begin{equation}
 Z_{\rm simple} =
\left(\frac{1}{\eta{(\tilde{q})}} \right)^2
\sum_{m,n}
\tilde{q}^{\frac{1}{2}(\frac{m^2}{gR^2}+n^2 g R^2)},
\end{equation}
with the coefficient unity.
Interpreted as the amplitude with two boundaries after
the folding, this means that the boundary entropy
is indeed zero.
Thus, there must be something wrong in
the assumptions led to eq.~\eqref{eq:wrongsimple}
(and to eq.~\eqref{eq:ZnG} in Ref.~\cite{Hsu-EE})
even for the free field.

\begin{figure}
\begin{center}
\includegraphics[width=0.6\textwidth]{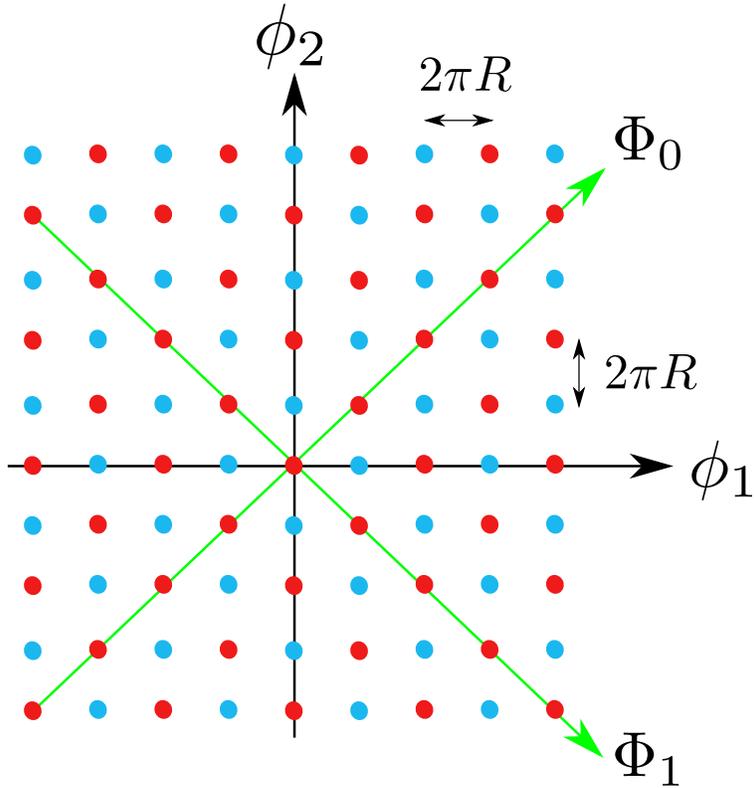}
\end{center}
\caption{The compactification lattice for two independent
boson fields $\phi_{1,2}$, each of which has compactification
radius $R$.
Points on the compactification lattice (shown as red and blue circles),
which is the square lattice with lattice constant $2\pi R$
in the $(\phi_1,\phi_2)$-plane, are identified.
The compactification is not imposed independently on
the linear combinations $\Phi_0=(\phi_1+\phi_2)/\sqrt{2}$
and $\Phi_1 = (\phi_1-\phi_2)/\sqrt{2}$.
The compactification lattice can be divided into two
sublattices (red and blue).
Considering only the red sublattice would be equivalent
to independent compactification of $\Phi_0$ and $\Phi_1$
with radius $\sqrt{2} R$.
However, the blue sublattice should also be included
in the compactification lattice.}
\label{fig:gluing}
\end{figure}

The problem was that the new fields were implicitly
assumed to obey the same compactification as the
original fields:
\begin{equation}
 \Phi_j \sim \Phi_j + 2 \pi R .
\end{equation}
In fact,
while the original fields are compactified independently,
a complication is introduced in the compactification
by a change of the basis.~\cite{WongAffleck}
This can been seen in Fig.~\ref{fig:gluing}.
The compactification in terms of the new fields $\Phi_{0,1}$
reads
\begin{align}
\Phi_0 & \sim \Phi_0 + 2 \pi n_0 \frac{R}{\sqrt{2}},\\
\Phi_1 & \sim \Phi_1 + 2 \pi n_1 \frac{R}{\sqrt{2}},
\end{align}
where
\begin{equation}
n_0 \equiv n_1 \mod{2}. 
\label{eq:gluing_n}
\end{equation}
Here $n_0 \equiv n_1 \equiv 0 \mod{2}$
and $n_0 \equiv n_1 \equiv 1 \mod{2}$ correspond
respectively to the red and blue sublattice
in Fig.~\ref{fig:gluing}.

We can also define the new fields $\Theta_{0,1}$ similarly
for the dual fields $\theta_{1,2}$.
Similarly to the case of $\Phi_{0,1}$, their compactification
is given as
\begin{align}
\Theta_0 & \sim \Theta_0 + 2 \pi m_0 \frac{1}{\sqrt{2}gR}, \\
\Theta_1 & \sim \Theta_1 + 2 \pi m_1 \frac{1}{\sqrt{2}gR},
\end{align}
where
\begin{equation}
m_0 \equiv m_1 \mod{2}. 
\label{eq:gluing_m}
\end{equation}

These considerations imply that the compactification
is not independent in terms of the new fields and
is subject to ``gluing conditions''~\eqref{eq:gluing_n}
and \eqref{eq:gluing_m} among different fields.
This aspect was ignored in Ref.~\cite{Hsu-EE}.
In the present simple example, we can see that
ignoring the gluing conditions leads to a wrong
equality~\eqref{eq:wrongsimple}.

This simple example demonstrates the importance
of the gluing conditions -- namely, that the linear combinations
of the compactified fields are not completely independent.
In fact, taking the gluing conditions into account,
the correct partition function can be reproduced
from the boundary CFT.
I describe this calculation in Appendix~\ref{app:simple}.

In passing, I note that in Ref.~\cite{WongAffleck}
additional gluing conditions between the dual
winding numbers ($n_j$ and $m_j$) were discussed.
These originate from Fermi statistics
of electrons in the microscopic model.
(See also Ref.~\cite{YJunction-full}.)
Those extra gluing conditions do not apply
to the present case.

\section{Cylinder geometry}
\label{sec:cylinder}

\begin{figure}
\begin{center}
\includegraphics[width=0.6\textwidth]{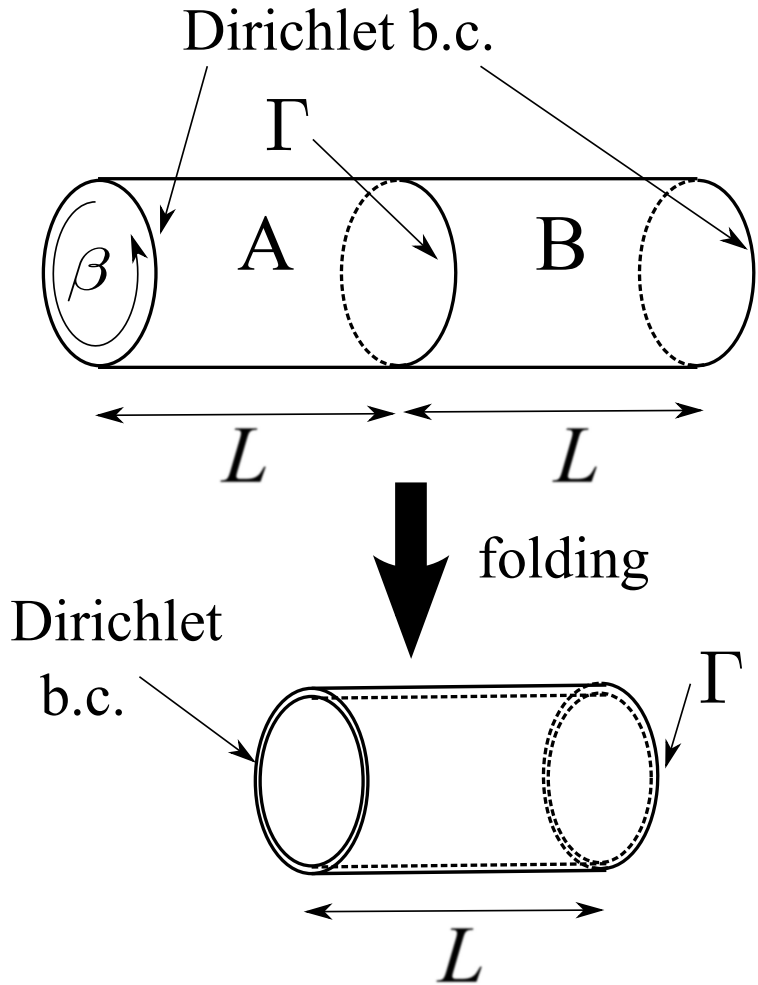}
\end{center}
\caption{The upper panel shows a cylinder
of circumference $\beta$ and length $2L$.
Dirichlet boundary condition $\phi=0$ is imposed
at the both ends.
The system is divided into two regions A and B with
length $L$ each, and I discuss the entanglement
entropy between the regions A and B.
In replica trick calculation, $n$-component free boson
field theory is defined on the cylinder.
It is folded onto a cylinder of length $L$, with
the open boundary condition on the one end.
The other end corresponds to the boundary $\Gamma$ between
the regions A and B.
After the folding, $\calN=2n$ component free boson is
defined on the cylinder.
}
\label{fig:cylinder}
\end{figure}

Now let us move on to the problem of the entanglement
entropy.
I start from the ``cylinder'' geometry introduced
in Ref.~\cite{Hsu-EE}, as shown in Fig.~\ref{fig:cylinder}.
Namely, I consider a cylinder of circumference $\beta$
and length $L_A + L_B$.
It is divided into two regions A and B, with length $L_A$ and
$L_B$ respectively.
For simplicity, I consider the case $L_A = L_B =L$.
At the two ends of the cylinder, I impose the
Dirichlet boundary condition $\phi=0$, as in Ref.~\cite{Hsu-EE}.

Following Refs.~\cite{FradkinMoore,Hsu-EE}, I employ replica trick.
At the boundary $\Gamma$ between A and B, the
condition~\eqref{eq:replica-bc} is imposed.
In order to apply boundary CFT to the present situation,
I invoke the folding technique
as introduced in Ref.~\cite{Isingdefect-NPB} and also
discussed in Sec.~\ref{sec:simple}.
Then the system may be regarded as a $2n$-component
free boson field theory on a cylinder of length $L$
and circumference $\beta$.
I simply label the fields after the folding
as $\phi_j$, where $j=1,2,\ldots, 2n$.
Here $\phi_j$ for $j>n$ represents the
``folding double'' of $\phi_{j-n}$.
I will denote the doubled number of components $2n$
as $\mathcal{N}$.
Each component obeys the compactification as in eq.~\eqref{eq:compact}.
Thus, the compactification lattice $\Lambda$ is the
$\mathcal{N}$-dimensional hypercubic lattice
with the lattice constant $R$.

At one end of the cylinder, which corresponds to
the two ends of the original cylinder before the folding,
the Dirichlet boundary condition
\begin{equation}
 \phi_j = 0 
\end{equation}
is imposed for $j=1,2, \ldots, \mathcal{N}$.

In calculating $Z_\calP$,
the condition~\eqref{eq:replica-bc} is imposed
at the other end, which corresponds to the boundary
between the two regions.
In terms of the $\mathcal{N}$-component field, it reads
\begin{equation}
\phi_1 = \phi_2 = \ldots = \phi_{\mathcal{N}-1} = \phi_{\mathcal{N}} .
\label{eq:replica-bc2}
\end{equation}
While
$\mathcal{N}=2n$ is an even integer
for the present application, the following construction
of the boundary state is valid for any (positive) integer
$\mathcal{N}$.

Following Refs.~\cite{FradkinMoore,Hsu-EE}, we may define the new basis by 
\begin{align}
\Phi_0 & \equiv \frac{1}{\sqrt{\mathcal{N}}}
 \sum_{j=1}^{\mathcal{N}} \phi_j 
\\
\Phi_1 & \equiv
\frac{1}{\sqrt{2}} \left( \phi_1 - \phi_2  \right)
\\
\vdots \notag \\
\Phi_{\mathcal{N}-1} & \equiv \ldots . \notag
\end{align}
We can also define their dual,
$\Theta_0, \ldots, \Theta_{\calN-1}$ by
linear combinations of $\theta_j$.
Among the new fields $\Phi_j$, only $\Phi_0$
obeys free (Neumann) boundary condition,
while all the others
$\Phi_1, \Phi_2, \ldots \Phi_{\mathcal{N}-1}$ is subject to
the Dirichlet boundary condition $\Phi_j = 0$.
Such a ``mixed'' (Dirichlet/Neumann) boundary condition has
been discussed in string theory~\cite{Ooguri-Oz-Yin} and
in condensed matter~\cite{YJunction-full} applications.

As we have seen in Sec.~\ref{sec:simple}
(which corresponds to $\mathcal{N}=2$),
the compactification is not independent
in terms of new fields $\Phi_j$.
It follows that the ``mixed''
boundary condition cannot be given just
by a simple product of Dirichlet/Neumann boundary states.
For a correct evaluation of the entanglement entropy
by the replica trick, we need to take
gluing conditions among $\calN = 2n$ component bosons
correctly into account.
However, it is cumbersome to keep track of
the gluing conditions explicitly as in Appendix~\ref{app:simple},
for larger number of fields.
Fortunately, the compactification of multi-component
boson field can be handled systematically
with a geometric formulation of the
``compactification lattice''.

I will denote the boundary state corresponding to
this ``replica'' boundary condition as $|\calP \rangle $.
The boundary condition implies that, the orthogonal matrix
$\calR$ in eq.~\eqref{eq:curr_bstate}
is reflection about the plane normal to
the $\mathcal{N}$-dimensional vector
\begin{equation} 
  \vec{d} \equiv (1,1,1, \ldots , 1)^T .
\end{equation}
Explicitly, the matrix is given as
\begin{equation}
 \calR = \mathbf{1} - 2 \vec{d} \vec{d}^T .
\label{eq:calRcalP}
\end{equation}

In order to construct the boundary state,
we need to identify the winding numbers which satisfy
eq.~\eqref{eq:allowed_zeromode}.
I define $\Xi_{D\calP}$ as the
intersection of $\Lambda^*/\sqrt{g}$
and the $\calN - 1$ dimensional hyperplane $\vec{d}\cdot\vec{K}=0$.
It is a Bravais lattice on the hyperplane $\vec{d}\cdot \vec{K}=0$.
The general solution $(\vec{R},\vec{K})$
of eq.~\eqref{eq:allowed_zeromode},
for the current choice of $\calR$ in eq.~\eqref{eq:calRcalP},
is given by
\begin{equation}
 \vec{R} = n_0 R \vec{d},
\label{eq:RincalP}
\end{equation}
for any integer $n_0$, and any $\vec{K} \in \Xi_{D\calP}$.

The corresponding boundary state may be written as
\begin{equation}
 | \calP (\vec{\theta}_0, \vec{\phi}_0)\rangle = 
g_\calP \sum_{\vec{R} = n_0 R \vec{d},\vec{K} \in \Xi_{D\calP}}
  e^{- i (\vec{\theta}_0 \cdot \vec{R}  + \vec{\phi}_0 \cdot \vec{K} )}
  |(\vec{R},\vec{K})\rangle \rangle,
\end{equation}
where the summation is taken over the solution 
of eq.~\eqref{eq:allowed_zeromode} discussed above.
$\vec{\theta}_0$ and $\vec{\phi}_0$ corresponds to the
boundary values of $\vec{\theta}$ and $\vec{\phi}$.
Since $\vec{R} \parallel \vec{d}$, 
components of $\vec{\theta}_0$ which are orthogonal
to $\vec{d}$ is irrelevant in this expression.
Thus we can assume without losing generality that
$\vec{\theta}_0$ is parallel to $\vec{d}$.
Similarly, components of $\vec{\phi}_0$ parallel
to $\vec{d}$ is irrelevant and thus $\vec{\phi}_0$
can be assumed to be orthogonal to $\vec{d}$.
These correspond to the fact that the present boundary
condition is Neumann on $\Phi_0$ (i.e. Dirichlet
on $\Theta_0$) and Dirichlet
on $\Phi_1,\Phi_2,\ldots,\Phi_{\calN-1}$.
$\vec{\theta}_0$ and $\vec{\phi}_0$ represent
the boundary value of $\Theta_0$ and $\Phi_{1,\cdots,\calN -1}$
respectively.

Actually, the ``replica'' boundary condition
implies all the ``difference'' fields $\Phi_{1,\cdots,\calN -1}$
vanish at the boundary, and
thus all components of $\vec{\phi}_0$ is zero for the present problem.
However it is useful to remember that this boundary
state is a special point in the continuous family
of boundary states labelled by $\vec{\theta}_0 \parallel \vec{d}$
and $\vec{\phi}_0 \perp \vec{d}$.
This demonstrates that $| \calP \rangle$ is the
boundary state indeed corresponding to
the Dirichlet boundary condition for
$\Theta_0$ and $\Phi_{1,2,\ldots,\calN-1}$.

The coefficient $g_\calP$ is determined by
Cardy's consistency condition.
The numerator of eq.~\eqref{eq:tr_rhoAn} in
this case is given by the amplitude
of the $\mathcal{N}$-component free boson field theory with
the boundary states
$|D\rangle$ and $|\calP\rangle$
at the two ends.
Only the winding number sectors common to
$|D\rangle$ and $|\calP\rangle$ contribute
to the amplitude.
Thus I find
\begin{equation}
 Z_{D\calP}(\tilde{q}) =
  g_D g_\calP \left( \frac{1}{\eta{(\tilde{q})}} \right)^{\mathcal{N}-1}
\tilde{q}^{-1/24}
\prod_{m=1}^\infty \frac{1}{1+\tilde{q}^m}
\sum_{\vec{K} \in \Xi_{D\calP}}  \tilde{q}^{\vec{K}^2/(4g)} .
\end{equation}
In order to satisfy Cardy's consistency condition upon
the modular transformation,
the amplitude must be written as
\begin{equation}
 Z_{D\calP}(\tilde{q}) =
z_{DN}(\tilde{q}) \calFt(\mathcal{N}-1;\Xi_{D\calP};\tilde{q}) ,
\end{equation}
where $\calFt$ is defined in eq.~\eqref{eq:Zb_closed}.
By comparison to eq.~\eqref{eq:Zb_closed}, I obtain
\begin{equation}
 g_D g_\calP = \frac{1}{\sqrt{2}} 2^{-(\mathcal{N}-1)/2} v_0(\Xi_{D\calP})
\end{equation}
In the present case,
\begin{equation}
 g_D = (2g)^{-\mathcal{N}/4} R^{-\mathcal{N}/2}
\end{equation}
which implies
\begin{equation}
  g_\calP = (\frac{g}{2})^{\mathcal{N}/4} R^{\mathcal{N}/2} v_0(\Xi_{D\calP})
\label{eq:gcalP_in_v0}
\end{equation}

\begin{figure}
\begin{center}
\includegraphics[width=0.6\textwidth]{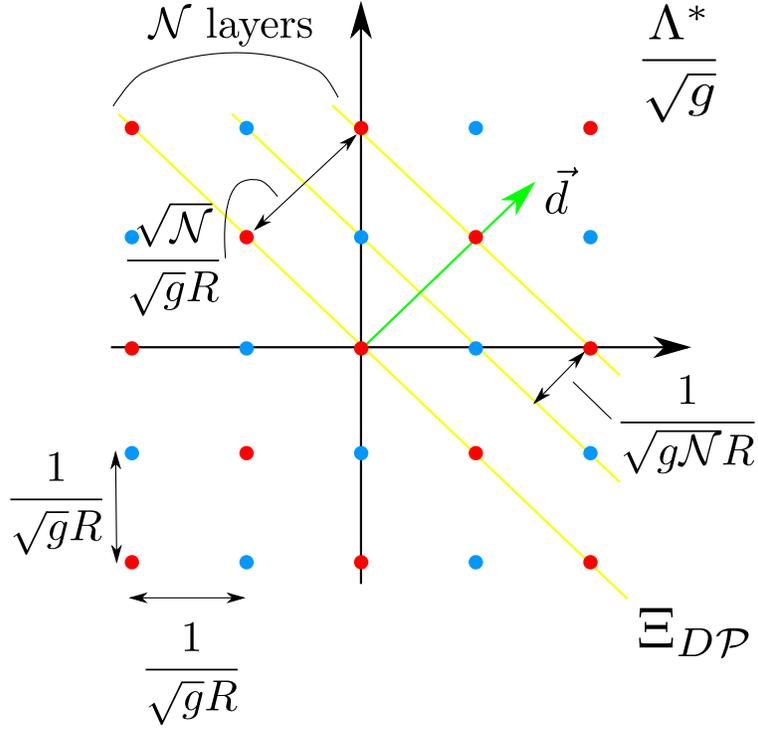}
\end{center}
\caption{
The construction of the lattice $\Xi_{D\calP}$ is
shown for $\calN=2$.
The lattice $\Lambda^*/\sqrt{g}$ is $\calN$-dimensional
hypercubic lattice (square lattice for $\calN =2$)
with lattice constant $1/\sqrt{g}R$, shown by red and
blue circles.
$\Xi_{D\calP}$ is given by the section of $\Lambda^*/\sqrt{g}$
by the $\calN -1$ dimensional hyperplane (line for $\calN =2$)
orthogonal to $\vec{d}=(1,1,\ldots,1)^T$, shown by
a yellow line.
$\Lambda^*/\sqrt{g}$ can be decomposed into copies of
$\Xi_{D\calP}$ displaced in parallel.
The distance between the neighboring copies of $\Xi_{D\calP}$
is given by $1/(\sqrt{g\calN} R)$.
}
\label{fig:section}
\end{figure}

Now I determine $v_0(\Xi_{D\calP})$, by calculating
$v_0(\frac{\Lambda^*}{\sqrt{g}})$ in two ways.
First, since $\Lambda^*$ is the $\mathcal{N}$-dimensional
hypercubic lattice with the lattice constant $1/R$,
we find
\begin{equation}
 v_0(\frac{\Lambda^*}{\sqrt{g}}) =
\left(\frac{1}{\sqrt{g}R}\right)^{\mathcal{N}} .
\label{eq:v0Lambdastar}
\end{equation}
On the other hand, $v_0(\frac{\Lambda^*}{\sqrt{g}})$ can be also
written in terms of $v_0(\Xi_{D\calP})$, as follows.
The $\mathcal{N}-1$ dimensional lattice $\Xi_{D\calP}$
is a intersection of $\Lambda^*$ and the hyperplane orthogonal
to $\vec{d}$.
This is illustrated in Fig.~\ref{fig:section} for the simplest
case of $\calN=2$.
The hypercubic lattice $\Lambda^*/\sqrt{g}$ can be decomposed into
parallel displacements of $\Xi_{D\calP}$.
Because $\Lambda^*$ is the simple cubic lattice with lattice
constant $1/R$,
for any vector $\vec{K} \in \Lambda^*/\sqrt{g}$,
\begin{equation}
\vec{K} \cdot \frac{\vec{d}}{\sqrt{\calN}} = \frac{m}{\sqrt{g \calN}R},
\label{eq:Kdist}
\end{equation}
with an integer $m$.
On the other hand, for any integer $m$,
there is always a vector $\vec{K} \in \Lambda^*/\sqrt{g}$
which satisfies eq.~\eqref{eq:Kdist}.
Eq.~\eqref{eq:Kdist} represents the
distance between the hyperplane and the origin.
When eq.~\eqref{eq:Kdist} holds,
$\vec{K}$ belongs to the $m$-th parallel displacement of $\Xi_{D\calP}$. 
Thus the distance between the neighboring hyperplanes
hosting a copy of $\Xi_{D\calP}$ is
\begin{equation}
 \frac{1}{\sqrt{g \mathcal{N}}R} ,
\end{equation}
and the volume of the unit cell of the original lattice
$\Lambda^*/\sqrt{g}$ is given as 
\begin{equation}
 v_0(\frac{\Lambda^*}{\sqrt{g}}) =
  \frac{1}{\sqrt{g \mathcal{N}}R}  v_0(\Xi_{D\calP}) .
\label{eq:v0Lambdastar2}
\end{equation}

By comparison of eqs.~\eqref{eq:v0Lambdastar}
and \eqref{eq:v0Lambdastar2}, I find
\begin{equation}
 v_0(\Xi_{D \calP}) =  \sqrt{2} g^{-(\mathcal{N}-1)/2}
  R^{-(\mathcal{N}-1)} \sqrt{\frac{\mathcal{N}}{2}}
\end{equation}
Therefore eq.~\eqref{eq:gcalP_in_v0} implies
\begin{align}
 g_\calP  &= (\sqrt{2g}R)^{-(\frac{\mathcal{N}}{2}-1)}
\sqrt{\frac{\mathcal{N}}{2}} \\
 & = (\sqrt{2g}R)^{-(n-1)} \sqrt{n}
\label{eq:gcalP}
\end{align}

The denominator of eq.~\eqref{eq:tr_rhoAn} corresponds to
the amplitude $Z_{DF}$ of the $\mathcal{N}$-component
free boson field theory with the boundary states
$|D\rangle$ and $| F \rangle$ at the two ends.
$|F \rangle$ is the boundary state corresponding to
\begin{equation}
 \phi_j = \phi_{j+n},  
\end{equation}
for $j = 1,2, \ldots ,n$.
Namely, each field is identified with its
``folding double'' at the boundary.
This corresponds to a boundary created by
folding $n$-component free boson in the bulk
without any defect.

As I have discussed in Sec.~\ref{sec:simple},
the boundary entropy for this artificially created ``boundary''
should be zero for any $n$.
Thus it must follow that
\begin{equation}
g_F=1 ,
\label{eq:gf_is_1}
\end{equation}
for any $n$.
In fact, eq.~\eqref{eq:gf_is_1}
for the boundary state $|F \rangle$ can be also shown
by an explicit calculation similar to that of $g_\calP$.
I note that, both $|F\rangle$ and $|\calP\rangle$
reduces to $|P\rangle$ for $n=1$ ($\calN =2$).

In the long cylinder limit $L \gg \beta$, $\tilde{q} \to 0$
and I obtain
\begin{align}
  Z_\calP = Z_{D\calP}(\tilde{q}) & \sim \tilde{q}^{-n/12} g_D g_\calP ,
\\
  Z_F = Z_{DF}(\tilde{q}) & \sim \tilde{q}^{-n/12} g_D g_F .
\end{align}
Thus, using eq.~\eqref{eq:gf_is_1}, I find
\begin{equation}
 \mathrm{Tr}{\rho_A}^n \sim g_\calP,
\end{equation}
in the long cylinder limit.
The von Neumann entanglement entropy from the replica trick
is thus given as
\begin{equation}
 S_E =  - \frac{\partial g_\calP}{\partial n} \big|_{n=1} .
\end{equation}
From eq.~\eqref{eq:gcalP}, for the quantum Lifshitz universality class,
\begin{equation}
 S_E = \log{\left( \sqrt{2g} R  \right)} - \frac{1}{2} .
\end{equation}
To match the convention in Refs.~\cite{Hsu-EE,Stephan-EE},
I take $g=1/2$ and thus $S_E = \log{R} - 1/2$.
This is different from the result reported
in Ref.~\cite{Hsu-EE}, by the second term $-1/2$.
On the other hand, it indeed agrees exactly
with that in Ref.~\cite{Stephan-EE} derived
by different approaches.
This shows that the basic ideas put forward in
Refs.~\cite{FradkinMoore,Hsu-EE} are correct, although care
must be taken in changing the basis.
Several results reported in Ref.~\cite{Hsu-EE}
are not valid owing to negligence of the subtlety.
The validity of the logarithmic term predicted in
Ref.~\cite{FradkinMoore} will be discussed in
Sec.~\ref{sec:conclusion}.

\section{Torus geometry}
\label{sec:torus}

In Ref.~\cite{Hsu-EE}, the entanglement entropy in
the torus geometry was also discussed.
Namely, a torus of total length $L_A+L_B$ is divided
into two regions A and B of cylindrical shape.
The boundary $\Gamma$
between A and B consists of two disjoint circles.

\begin{figure}
\begin{center}
\includegraphics[width=0.6\textwidth]{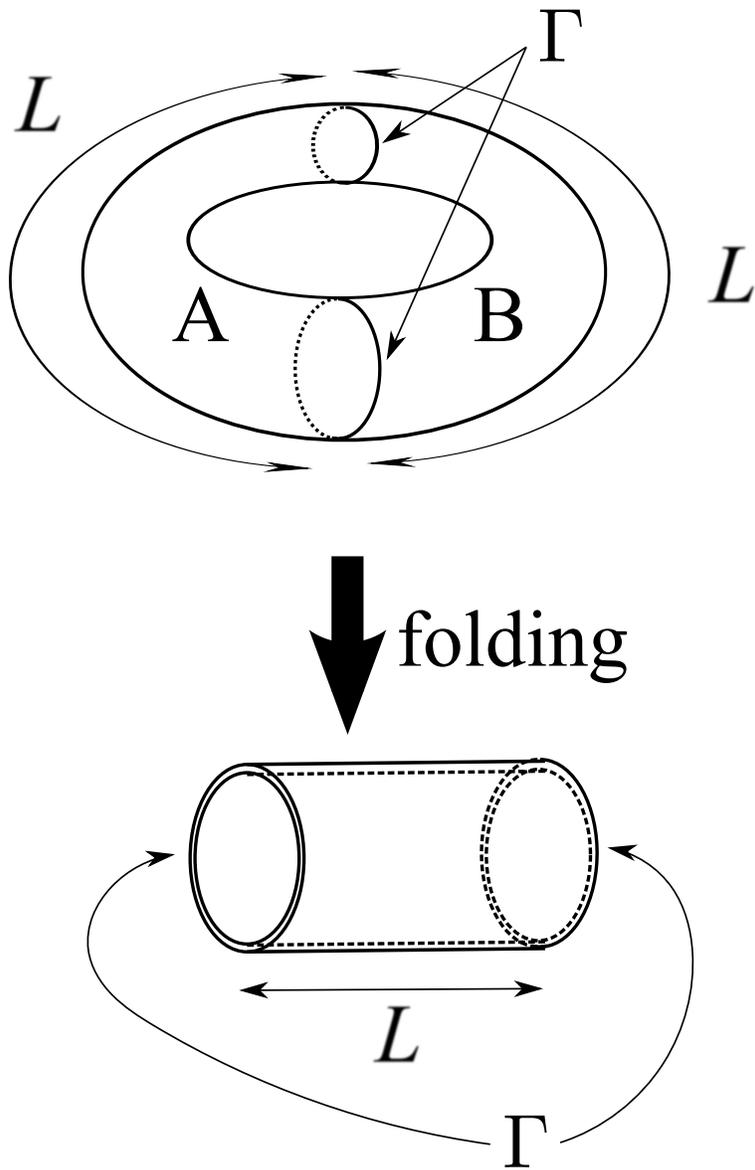} 
\end{center}
\caption{The torus (upper panel)
of size $\beta \times 2L$
is divided into two regions A and B with
length $L$ each, and I discuss the entanglement
entropy between the regions A and B.
Folding reduces the torus to the cylinder of length $L$,
where the boundaries at the two ends
correspond to the boundary $\Gamma$
between regions A and B.
}
\label{fig:torus}
\end{figure}

Again, for simplicity I consider the case $L_A = L_B =L$
and apply the folding trick to reduce the problem
to $\calN=2n$-component free boson field theory on a cylinder
of length $L$, as in Fig.~\ref{fig:torus}.
By construction, in the present case,
the ``replica'' boundary condition $\calP$
should be imposed on the both boundaries,
in order to calculate the numerator of
eq.~\eqref{eq:tr_rhoAn}.
This is contrasted to the case of cylinder geometry
discussed in Sec.~\ref{sec:cylinder}, where
the replica boundary condition is imposed at one end
and the Dirichlet boundary condition at the other.

Thus I find
\begin{equation}
 \mathrm{Tr} {\rho_A}^n =
 \frac{Z_{\calP \calP}(\tilde{q})}{Z_{FF}(\tilde{q})} .
\end{equation} 
In the long cylinder limit,
\begin{equation}
 \mathrm{Tr} {\rho_A}^n \sim  \left(\frac{g_\calP}{g_F}\right)^2
 = {g_\calP}^2 .
\end{equation}
This implies that
\begin{equation}
 S^{\mbox{\scriptsize (torus)}}_E =
 2 S^{\mbox{\scriptsize (cylinder)}}_E,
\label{eq:SE_torus}
\end{equation}
which leads to
\begin{equation}
 S^{\mbox{\scriptsize (torus)}}_E =
 2 \left( \log{\sqrt{2g}R} - \frac{1}{2} \right) .
\end{equation}
This result is, again, different from that in Ref.~\cite{Hsu-EE}.
In particular eq.~\eqref{eq:SE_torus} does not hold
in Ref.~\cite{Hsu-EE}.
On the other hand, eq.~\eqref{eq:SE_torus} is a general
consequence of the boundary CFT, independent of
the value of $S_E$.
In fact, although it was not explicitly discussed in
Ref.~\cite{Stephan-EE}, eq.~\eqref{eq:SE_torus}
is also a general consequence of the mapping to
classical statistical problem used in Ref.~\cite{Stephan-EE}.
Thus the violation of the relation~\eqref{eq:SE_torus}
is a clear signature of the problem in the calculation in
Ref.~\cite{Hsu-EE}.

I emphasize that eq.~\eqref{eq:SE_torus} does \emph{not} hold
for the entanglement entropy in  general systems.
It only applies to the critical wavefunction
described as in eq.~\eqref{eq:wfCFT},
for which the entanglement entropy can be related
to classical (Shannon) entropy.~\cite{Stephan-EE}
In fact, eq.~\eqref{eq:SE_torus} is violated in
topological $Z_2$ spin liquid phase.

\section{Conclusion and Discussions}
\label{sec:conclusion}

I have discussed the entanglement entropy
in two-dimensional conformal critical points,
in particular those described by
free boson CFT
(quantum Lifshitz universality class).

Calculations in Refs.~\cite{FradkinMoore,Hsu-EE} 
were based on the fundamental formula~\eqref{eq:ZnG}
(or equivalently eq.~\eqref{eq:SEinFM}).
However, there are two problems in this formula.
First, in a general interacting theory,
linear combinations of the original fields are not
independent of each other
as it was implicitly assumed
in Refs.~\cite{FradkinMoore,Hsu-EE}.
On the other hand, CFTs other than
the free boson field theory also
admit free field representations.
It might be used to extend the approach to
general CFTs.
However, it would be a nontrivial problem which
requires further careful investigations.

Second, for the free boson field theory (which describes
the quantum Lifshitz universality class), the bulk
interaction is absent and the linear combinations
appear independent.
However, even in this case, they are not completely
independent because of the intertwined compactification.
I have demonstrated the importance of the
nontrivial compactification using the simple
example of single-component free boson field on
a torus, which can be regarded as two-component
free boson field with boundaries.
``Mixed'' Dirichlet/Neumann boundary conditions
for general number of components, which
appears in the replica trick calculation of
the entanglement entropy, can be handled
with a geometric formulation based on
the compactification lattice.
The constant part in the entanglement entropy
corresponds to the universal boundary entropy
in the CFT.
This supports the universality of the constant
term as proposed in Ref.~\cite{Hsu-EE}.
In fact, the entanglement entropy obtained
for the quantum Lifshitz universality class
with the boundary CFT
agrees exactly with that obtained with
different methods in Ref.~\cite{Stephan-EE}.
An incorrect value was reported in Ref.~\cite{Hsu-EE}
because of the compactification was not properly
taken into account.

The predictions made in Ref.~\cite{FradkinMoore}
on the logarithmic term in eq.~\eqref{eq:SEwithlog}
should also be re-examined, since
the fundamental formula~\eqref{eq:ZnG}, on which their derivation
is based, does not hold as an exact identity.
Nevertheless, their prediction on the logarithmic term
could still stand valid for the following reason,
in particular for the free boson (quantum Lifshitz) case.
For the free boson field theory, in Ref.~\cite{FradkinMoore},
they found the logarithmic term independent of the compactification
radius.
This suggests that the logarithmic term could be attributed
solely to oscillator modes.
For oscillator mode contributions, there is no subtlety
discussed in the present paper and the ``changing the basis''
trick could be justified.
From this perspective, it seems quite possible that
their prediction on the logarithmic term is correct
despite the subtle problem with eq.~\eqref{eq:ZnG}.
It would mean that eq.~\eqref{eq:ZnG} is valid in some
restricted sense for determining the logarithmic term,
although it is certainly not valid for determining
the universal constant term.
For general CFTs other than free boson, there is more
problem in the ``changing the basis'' trick due to the
bulk interaction. It might be still possible that
the logarithmic term depends only on the central charge
and the prediction in Ref.~\cite{FradkinMoore} is also
valid for general CFTs, but it seems less convincing
than in the free boson case, at this point.

In any case, in the present paper, I do not have
any concrete result concerning the logarithmic term,
and cannot draw a definitive conclusion about the
prediction in Ref.~\cite{FradkinMoore} on the logarithmic
term. I hope that the present paper will stimulate
further progress in understanding of the logarithmic term.

It should be noted that the present paper, as well
as the original work~\cite{Hsu-EE}, is entirely
based on (a simple implementation of) the replica trick.
Its validity is by no means obvious.
For the present case of free boson CFT,
the agreement with different methods~\cite{Stephan-EE}
implies that it is indeed valid.
However, recently, its breakdown is suggested
when the CFT corresponds to
critical Ising model.~\cite{Stephan-Ising}
A general understanding of the issue is
an important open problem.

Finally, when this paper was close to completion,
a paper by Hsu and Fradkin has appeared~\cite{HsuFradkin}.
There, they did not rely on eq.~\eqref{eq:SEinFM}, which
is not valid as I have discussed.
Instead, they attempted to construct the boundary state
in the spirit similar to the present paper.
However, although their construction
(eq.~(14), or eq.~(18) with eq.~(19) of Ref.~\cite{HsuFradkin})
should give a consistent boundary state,
it does not correspond to the required boundary
condition~\eqref{eq:replica-bc2} for the problem.
As an indication, the condition~\eqref{eq:replica-bc2}
is invariant under any permutation of boson fields $\phi_j$
while their construction is not.

Furthermore, apparently there are several confusions
in Ref.~\cite{HsuFradkin}.
For example, in the original paper~\cite{Hsu-EE},
as well as in Ref.~\cite{Stephan-EE} and in the present
paper, the universal constant in the entanglement entropy
was derived in the ``long cylinder limit'' $L \gg \beta$
(in the notation of the present paper).
However, in Ref.~\cite{Hsu-EE}, the definition of the
lengths is somehow exchanged and the opposite limit is taken.
As discussed in the present paper, in the long cylinder limit
as introduced originally in Ref.~\cite{Hsu-EE},
the universal constant part of the entanglement entropy
should correspond to the universal
boundary entropy (exponential of the groundstate degeneracy)
in the boundary CFT.

\section*{Acknowledgements}

I am grateful to Gr\'{e}goire Misguich for
introducing the problem to me, and for comments on the paper.
I also thank Fabien Alet, Claudio Castelnovo, Michael Freedman,
Shunsuke Furukawa, Joel Moore, Hirosi Ooguri, and
Tadashi Takayanagi for useful discussions.
This work is initiated at
Laboratoire de Physique Th\'{e}orique at IRSAMC Toulouse,
and partially carried out at the
International Workshop on Topological Order and Quantum Computation
at Richard B. Gump South Pacific Research Station of UC Bekeley,
and at the Summer Workshop ``Low Dimensional Topological Matter''
at Aspen Center for Physics.

This work is supported in part by Grant-in-Aid for
Scientific Research on Innovative Areas
No. 20102008 from MEXT of Japan, 
and by Grant-in-Aid for Challenging Exploratory Research
No. 20654030 from JSPS.

\appendix

\section{Boundary CFT of multicomponent boson field}
\label{app:multiboson}

In this Appendix, I summarize the relevant formulae in
the boundary CFT of multicomponent
compactified free boson field theory.

The boundary CFT was largely developed by
Cardy~\cite{Cardy-BCFT1984,Cardy-BCFT1989}.
The development of boundary CFT of free boson field
theory was also started in the context of
string theory~\cite{Polchinski-Cai,Callan-LNY},
and continued for example
in Refs.~\cite{Callan-KLM-BCFT,Callan-KMY-Magnetic,Yegulalp1995,Ooguri-Oz-Yin}.
For a review in the string theory context,
see Refs.~\cite{DiVecchia-Liccardo-DbraneRev}.

On the other hand, the boundary CFT
is also relevant for some problems in condensed matter
physics, such as quantum impurity problems and
junction of one-dimensional quantum systems.
The boundary CFT of free boson field
theory is applied to impurities in quantum spin chains
by Eggert and Affleck~\cite{EggertAffleck-PRB1992}.
The boundary CFT of multicomponent free boson
is applied to impurities in quantum wires
by Wong and Affleck~\cite{WongAffleck},
and further discussed
in related problems~\cite{QBM,YJunction-PRL,YJunction-full}

The multicomponent free boson field theory is
defined by the Lagrangian density
\begin{equation}
 \calL = \frac{g}{4 \pi}(\partial_\mu \vec{\phi})^2,
\label{eq:multiLag}
\end{equation}
where $\vec{\phi}$ is a $\calN$-dimensional vector.

I introduce the multidimensional generalization of
the boson compactification as
\begin{equation}
 \vec{\phi} \sim \vec{\phi} + 2 \pi \vec{R},
\label{eq:multicompact}
\end{equation}
where $\vec{R} \in \Lambda$ for a Bravais lattice
$\Lambda$, which is called as compactification lattice.
For $\calN$ independent copies of the single-component
boson compactified as in eq.~\eqref{eq:compact},
the compactification lattice $\Lambda$ is simply
a hypercubic lattice with lattice constant $R$.
However, it is useful to formulate allowing more general
compactification lattice.

I note that, as in the single-component case,
the coupling constant $g$ can be set to any value
by renormalizing the field $\vec{\phi}$,
which also renormalize the compactification lattice.
Thus $g$ is a redundant parameter once I consider
general compactification lattice.
Nevertheless, I keep the coupling constant $g$
for purpose of comparison as I have discussed
for the single-component case.
On the other hand, for general compactifications,
it is impossible to fix the compactification lattice
by renormalizing $\vec{\phi}$ within the
Lagrangian density of the form~\eqref{eq:multiLag}.

If we define the theory~\eqref{eq:multiLag} on a finite
length $\beta$ with the periodic boundary condition,
the canonically quantized operator $\vec{\phi}$ is
given by the mode expansion
\begin{equation}
\vec{\phi} (t,x)=
{\vec{\phi}}^{(0)} +
\frac{2\pi}{\beta}\left[\vec{R}x+ \vec{P} t\right]+
\frac{1}{\sqrt{2g}}\sum_{n=1}^\infty \frac{1}{\sqrt{n}}\left\{ 
\vec{a}_n^L\exp{\left[-inx_+\frac{2\pi}{\beta}\right]}+
\vec{a}_n^R\exp{\left[-inx_-\frac{2\pi}{\beta}\right]}+
\mbox{h.c.} \right\},
\label{eq:phi_modes1}
\end{equation}
where $x_\pm \equiv x \pm t$.
$\vec{R} \in \Lambda$ represents the winding number
of $\vec{\phi}$ when $x$ goes around the system.
The canonical commutation relation of $\vec{\phi}$
with the conjugate momentum field implies
\begin{equation}
 [ \phi^{(0)}_j, P_k]  = i \delta_{jk} .
\label{eq:phiPcr}
\end{equation}
Since the constant part $\vec{\phi}^{(0)}$ is also
subject to the compactification as in
eq.~\eqref{eq:multicompact},
the eigenvalues of the conjugate ``momentum'' operator
$\vec{P}$ is quantized as
\begin{equation}
 \vec{P} = \frac{1}{g} \vec{K},
\label{eq:PbyK}
\end{equation}
where $\vec{K}$ belongs to the dual lattice
$\Lambda^*$, which is defined as a set of
all $\vec{K}$'s which satisfies
\begin{equation}
 \vec{K} \cdot \vec{R} = \mbox{integer}, 
\end{equation}
for any $\vec{R} \in \Lambda$.
The momentum operator $\vec{P}$ in
eq.~\eqref{eq:phi_modes1} is often rewritten using
eq.~\eqref{eq:PbyK}.
However, it should be kept in mind that
$\vec{K}$ represents the eigenvalue of $\vec{P}$
as in eq.~\eqref{eq:PbyK}.

The boson field $\vec{\phi}$ can be decomposed into
left-moving and right-moving components as
\begin{equation}
 \vec{\phi} = \vec{\phi}^L(x_+) + \vec{\phi}^R(x_-) .
\end{equation} 
We can introduce the dual boson field as
\begin{equation}
 \vec{\theta} \equiv g(\vec{\phi}^L - \vec{\phi}^R) .
\end{equation}
From eq.~\eqref{eq:phi_modes1},
mode expansion of $\vec{\theta}$ is given as
\begin{equation}
\vec{\theta} (t,x)=
\vec{\theta}^{(0)} +
\frac{2\pi}{\beta} \left[\vec{K}x + g \vec{R} t\right]+
{\sqrt{\frac{g}{2}}}\sum_{n=1}^\infty \frac{1}{\sqrt{n}}\left\{ 
\vec{a}_n^L\exp{\left[-inx_+\frac{2\pi}{\beta}\right]}+
\vec{a}_n^R\exp{\left[-inx_-\frac{2\pi}{\beta}\right]}+
\mbox{h.c.} \right\} .
\label{eq:theta_modes}
\end{equation}
In this expression, the roles of $\vec{K} \in \Lambda^*$
and $g \vec{R} \in g \Lambda$ are interchanged compared
to the mode expansion of $\vec{\phi}$ in
eq.~\eqref{eq:phi_modes1}.
This implies that the dual field obeys compactification
\begin{equation}
 \vec{\theta} \sim \vec{\theta} + \frac{2\pi}{g} \vec{K},
\end{equation}
where $\vec{K} \in \Lambda^*$.
$g \hat{R}$ is now interpreted as the eigenvalue of
the dual ``momentum'' operator, which obeys
canonical commutation relation similar to
eq.~\eqref{eq:phiPcr} with the operator
$\vec{\theta}^{(0)}$.

The Lagrangian density~\eqref{eq:multiLag} can be
also written in terms of $\vec{\theta}$ as
\begin{equation}
 \calL = \frac{1}{4 \pi g}(\partial_\mu \vec{\theta})^2.
\end{equation}
Although $\vec{\phi}$ and $\vec{\theta}$ are
mutually non-local, construction of a complete set
of physical operators requires both $\vec{\phi}$
and $\vec{\theta}$.

When we take boundary of a $1+1$ dimensional field
theory orthogonal to the (imaginary) time axis,
the boundary condition can be represented by
the initial (or final) state.
This is called as boundary state and is useful in
systematic study of boundary conditions.
In fact, nontrivial boundary conditions often cannot
be defined precisely without introducing
the boundary states.
When the boundary condition is conformally invariant,
the corresponding boundary state $ B\rangle$ satisfies
\begin{equation}
 \left( L_m - \bar{L}_{-m}\right) |B\rangle = 0,
\label{eq:conf_bstate}
\end{equation}
for any integer $m$, where $L_m$ are generators of
the Virasoro algebra.
For the multicomponent boson field, the Virasoro generators
are given as
\begin{align}
L_m &= \frac{1}{2} \sum_l : \vec{\alpha}^L_{m-l} \vec{\alpha}^L_l : ,
\\
\bar{L}_m &=
\frac{1}{2} \sum_l : \vec{\alpha}^R_{m-l} \vec{\alpha}^R_l : ,
\end{align}
where 
\begin{align}
 \vec{\alpha}^L_n  = \left\{
\begin{array}{cc}
-i \sqrt{n} \vec{a}^L_n  & (n>0)
\\
\frac{1}{\sqrt{2}}
\left( \sqrt{g}\vec{R} + \frac{1}{\sqrt{g}} \vec{K} \right)
& (n=0)
\\
i \sqrt{n} (\vec{a}^L_{-n})^\dagger  & (n<0)
\end{array}
\right. , & \;\;
 \vec{\alpha}^R_n = \left\{
\begin{array}{cc}
-i \sqrt{n} \vec{a}^R_n  & (n>0)
\\
\frac{1}{\sqrt{2}}
\left( - \sqrt{g} \vec{R} + \frac{1}{\sqrt{g}} \vec{K} \right)
& (n=0)
\\
i \sqrt{n} (\vec{a}^R_{-n})^\dagger  & (n<0)
\end{array}
\right. .
\end{align}
Here each component of $\vec{a}^{L,R}_n$
is a boson annihilation operator, 
$\vec{R} \in \Lambda$, and $\vec{K} \in \Lambda^*$.
$\vec{a}^{L,R}_n$ and their Hermitian conjugates (boson creation
operators) represent the oscillator modes, corresponding
to the quantized normal mode oscillations.
These oscillator modes do not depend on the compactification.
On the other hand, $\vec{\alpha}^{L,R}$ represent
zero modes which are affected by the compactification.

The general solution of the conformal invariant
boundary state~\eqref{eq:conf_bstate}  for the
multicomponent free boson is not known.
However, a sufficient condition for eq.~\eqref{eq:conf_bstate} 
can be given as
\begin{equation}
 \left( \vec{\alpha}^L_m - \calR \vec{\alpha}^R_{-m} \right)|B\rangle = 0,
\label{eq:curr_bstate}
\end{equation}
for arbitrary integer $m$.
Here $\calR$ is an $\calN \times \calN$
orthogonal matrix independent of $m$.
Eq.~\eqref{eq:curr_bstate} for $m \neq 0$ determines
the boundary state to be of the form
\begin{equation}
 \exp{\left( -
\sum_{n=1}^\infty ({{\vec{a}}^L}_n)^\dagger \calR (\vec{a}^R_n)^\dagger
\right)} | \mbox{vac} \rangle ,
\label{eq:boson_Ishibashi}
\end{equation}
where $|\mbox{vac}\rangle$ is an oscillator vacuum.
This is a free boson version of Ishibashi state~\cite{Ishibashi-state},
which is conformal invariant.

In fact, there is an infinite number of oscillator vacua
characterized by the zero mode quantum numbers.
Following Ref.~\cite{YJunction-full}, I label these vacua as
\begin{equation}
 | (\vec{R}, \vec{K}) \rangle.
\label{eq:vacRK}
\end{equation}
$\vec{R}$ and $\vec{K}$ can be interpreted
as the ``winding numbers'' of $\vec{\varphi}$
and of its dual $\vec{\theta}$, respectively,
along the boundary.
The Ishibashi state
obtained as eq.~\eqref{eq:boson_Ishibashi} from
eq.~\eqref{eq:vacRK} is denoted as
\begin{equation}
| (\vec{R}, \vec{K}) \rangle \rangle .
\end{equation}

Eq.~\eqref{eq:curr_bstate} for $n=0$ puts a restriction
on the vacua which can appear in the boundary state.
That is, the quantum numbers must satisfy
\begin{equation}
 \left( \sqrt{g} \vec{R} + \frac{1}{\sqrt{g}} \vec{K}\right) =
\calR
 \left( - \sqrt{g} \vec{R} + \frac{1}{\sqrt{g}} \vec{K}\right).
\label{eq:allowed_zeromode}
\end{equation}
For given orthogonal matrix $\calR$, generally there
is an infinite number of solutions $(\vec{R}, \vec{K})$
which satisfy this requirement.
Any linear combination of the Ishibashi states
built from these vacua satisfies
the conformal invariance~\eqref{eq:conf_bstate}.

However, a physical boundary state must also
satisfy Cardy's consistency condition, stated as follows.
For a pair of given boundary conditions A and B,
we can define the amplitude (partition function)
\begin{equation}
 Z_{\calA \calB}(\tilde{q}) =  \langle \calA |
e^{-L \hat{H}_{P}} | \calB \rangle
\end{equation}
where $\tilde{q}=...$
By modular transformation, we can express this
amplitude as a function of $q=e^{}$,
\begin{equation}
 Z_{\calA \calB} (q) = \sum_h N_{\calA \calB}^h \chi^{\mathrm{Vir}}_h (q) 
\end{equation}
where $\chi^{\mathrm{Vir}}_h(q)$ is a character
of the Virasoro algebra.
Since $N_{\calA \calB}^h$ can be interpreted as the
number of primary fields with conformal weight $h$,
it has to be a non-negative integer.
This is Cardy's consistency condition.
Usually it is also required that $N_{\calA \calA}^0=1$,
where $h=0$ corresponds to the identity operator.
If one takes just a single Ishibashi state,
Cardy's condition cannot be satisfied.
Generally, we take linear combination of
Ishibashi states for all the zero modes allowed
by eq.~\eqref{eq:allowed_zeromode}.
It is also possible to construct consistent boundary states
using only subset of zero modes allowed by
eq.~\eqref{eq:allowed_zeromode}.
However, those boundary states are more unstable
and the most stable boundary states for a given $\calR$
turns out to be linear combinations of all the allowed
zero mode vacua~\cite{YJunction-full}.

For example, taking $\calR = 1$ gives Dirichlet boundary state.
The solution of eq.~\eqref{eq:allowed_zeromode} for $\calR=1$
is given by $\vec{R}=0$.
Thus the Dirichlet boundary state is given as
\begin{equation}
 |D(\vec{\phi}_0) \rangle =
 g_D \sum_{\vec{K} \in \Lambda^*}
e^{- i \vec{\phi}_0 \cdot \vec{K}}
|(\vec{0},\vec{K}) \rangle \rangle, 
\label{eq:Dstate_phi0}
\end{equation}
where
the summation in $\vec{K}$ is taken over the entire
dual lattice $\Lambda^*$,
and $\vec{\phi}_0$ is a constant $\calN$-dimensional vector.
(There is an unfortunate conflict of notation; $g_D$ here
is the coefficient of the boundary state and is a completely
different quantity from the coupling constant $g$
defined in eq.~\eqref{eq:Lag}.)

Physically, the constant vector $\vec{\phi}_0$ corresponds to the
boundary value of the field $\vec{\phi}$.
This can be seen as follows.
Let us consider the operator
\begin{equation}
 e^{  i  \vec{\phi} \cdot \vec{K}_0} 
\end{equation}
where $\vec{K}_0$ is a constant vector belonging to the dual lattice
$\Lambda^*$.
This is a unique valued operator under the
compactification~\ref{eq:multicompact}.
We apply this operator to the boundary
state~\eqref{eq:Dstate_phi0} at (imaginary) time $0$.
With respect to oscillator modes,
each Ishibashi state $|(\vec{R},\vec{K})\rangle \rangle$
constructed as in eq.~\eqref{eq:boson_Ishibashi}
is a kind of coherent state.
We observe that, for general $\calR$,
\begin{equation}
 \vec{a}^L_n | (\vec{R},\vec{K}) \rangle \rangle
  = - \calR (\vec{a}^R_n)^{\dagger} | (\vec{R},\vec{K}) \rangle \rangle.
\end{equation}
For the Dirichlet boundary condition, $\calR=1$ and thus 
applying the oscillator part of $\vec{\phi}$ to
the Ishibashi state yields zero.
The winding number $\vec{R}$ is zero in all the Ishibashi
states in the boundary state~\eqref{eq:Dstate_phi0}, and
at $t=0$, contribution of the ``momentum'' eigenvalue
$\vec{K}$ to $\vec{\phi}$ vanishes.
Thus we find
\begin{equation}
  e^{  i  \vec{\phi}(x,0) \cdot \vec{K}_0} |D (\vec{\phi}_0) \rangle
=
e^{  i  \vec{\phi}^{(0)} \cdot \vec{K}_0} |D (\vec{\phi}_0) \rangle.
\label{eq:expphi_D1}
\end{equation}
Now the commutation relation~\eqref{eq:phiPcr} 
implies
\begin{equation}
 e^{i  \vec{\phi}^{(0)} \cdot \vec{K}_0 } |(\vec{R},\vec{K}) \rangle
= | (\vec{R},\vec{K}+\vec{K}_0) \rangle.
\end{equation}
Combining this with eq.~\eqref{eq:expphi_D1}, we find the
eigenequation
\begin{equation}
  e^{  i  \vec{\phi}(x,0) \cdot \vec{K}_0} |D (\vec{\phi}_0) \rangle
 = e^{i \vec{\phi}_0 \cdot \vec{K}_0 } |D (\vec{\phi}_0) \rangle .
\end{equation}
This proves that $\vec{\phi}_0$ can indeed be interpreted
as the boundary value of the field $\vec{\phi}$.
The Dirichlet boundary state is actually a continuous
family of boundary states parametrized by $\vec{\phi}_0$.
By symmetry, physical properties such as
boundary entropy and scaling dimensions
of the boundary operators are independent of $\vec{\phi}_0$.
Thus the boundary value is often set to zero for simplicity.
I denote $|D(\vec{\phi}_0=\vec{0})\rangle$ simply
by $|D\rangle$.

The Dirichlet-Dirichlet amplitude is given as
\begin{equation}
 Z_{DD}(\tilde{q}) =
\calFt(\calN;\frac{\Lambda^*}{\sqrt{g}};\tilde{q}),
\end{equation}
where 
\begin{align}
\calFt(\calN;\Xi;\tilde{q}) & \equiv
2^{-\calN/2} v_0(\Xi) \left( \frac{1}{\eta{(\tilde{q})}}\right)^\calN
\sum_{\vec{v} \in \Xi} \tilde{q}^{\vec{v}^2/4}
\label{eq:Zb_closed}
\\
= \calF(\calN; \Xi^* ; q)
&=
\left( \frac{1}{\eta{(q)}} \right)^c
\sum_{\vec{u} \in \Xi^*} q^{\vec{u}^2} .
\label{eq:Zb_open}
\end{align}
Here $\Xi$ is a Bravais lattice and $\Xi^*$
is its dual,  $v_0(\Xi)$ is the
volume of the unit cell of the lattice $\Xi$,
and
\begin{equation}
 \eta{(q)} \equiv q^{1/24} \prod_{n=1}^{\infty} (1-q^n),
\label{eq:Dedekind_eta}
\end{equation}
is the Dedekind eta function.
I note that
\begin{equation}
 v_0(\Xi^*) = \frac{1}{v_0(\Xi)} .
\end{equation}
Equality between eq.~\eqref{eq:Zb_closed} and
eq.~\eqref{eq:Zb_open} holds thanks to
the multidimensional generalization
of Poisson summation formula.

The coefficient in eq.~\eqref{eq:Zb_closed}
is fixed so that
the amplitude in the form~\eqref{eq:Zb_open}
satisfies Cardy's consistency condition.

By comparison to eq.~\eqref{eq:Zb_closed},
we find
\begin{equation}
 g_D = (2g)^{-\calN/4} (v_0(\Lambda))^{-1/2},
\end{equation}
where $v_0(\Lambda)$ is the volume of the unit cell
of the compactification lattice $\Lambda$.
$g_D$ can be interpreted as universal non-integer
groundstate degeneracy; in other words,
$\log{g_D}$ is the boundary entropy for the Dirichlet
boundary state.

Likewise, the Neumann boundary condition
corresponds to $\calR=-1$.
The Neumann boundary condition for $\vec{\phi}$
is equivalent to the Dirichlet boundary condition
for the dual field $\vec{\theta}$.
The Neumann boundary state is given as
\begin{equation}
 |N (\vec{\theta}_0)\rangle =
 g_N \sum_{\vec{R} \in \Lambda}
 e^{- i 2 \pi \vec{\theta}_0 \cdot \vec{R}}
|(\vec{R},\vec{0}) \rangle \rangle,
\end{equation}
where $\vec{\theta}_0$ is the boundary value of
the dual field $\vec{\theta}$.
The Neumann-Neumann amplitude reads
\begin{equation}
 Z_{NN}(\tilde{q}) = \calFt(\calN;\sqrt{g} \Lambda; \tilde{q}) .
\end{equation}
It follows that the groundstate degeneracy is
\begin{equation}
 g_N = \left(\frac{g}{2} \right)^{\calN/4} (v_0(\Lambda))^{1/2} .
\end{equation}

The Dirichlet-Neumann amplitude is also of interest;
it must satisfy Cardy's consistency condition as well.
I note that, the Hamiltonian time evolution does not
change the winding numbers from those in the initial state.
Thus the only oscillator vacuum which contributes to the
amplitude is $|(\vec{0},\vec{0})\rangle$, and
no summation over zero modes appears in the Dirichlet-Neumann
amplitude.
The amplitude is thus given only by oscillator mode contributions as
\begin{equation}
 Z_{ND}(\tilde{q}) = \left( z_{ND}(\tilde{q}) \right)^\calN,
\end{equation}
where
\begin{align}
 z_{ND}(\tilde{q}) = &
\frac{1}{\sqrt{2}} \tilde{q}^{-1/24}
\prod_{n=1}^\infty \frac{1}{1+\tilde{q}^n}
=
\frac{1}{\sqrt{2} \eta{(\tilde{q})}} \vartheta_4(\tilde{q}^2)
\notag \\
=& 
\frac{1}{2 \eta{(q)}} \vartheta_2 (q^{1/2}),
\end{align}
is the Dirichlet-Neumann amplitude for a single component boson. 
Here $\vartheta_{2,4}$ are Jacobi's theta function defined as
\begin{align}
\vartheta_2(q) & \equiv \sum_{n = -\infty}^{\infty} q^{(n+1/2)^2} ,\\
\vartheta_4(q) & \equiv \sum_{n = -\infty}^{\infty} (-1)^n q^{n^2}.
\end{align}
This amplitude indeed satisfies Cardy's consistency condition,
as it should.
It is also noted that this amplitude does not
depend on the compactification.

\section{Construction of the boundary state for the
simple example based on the gluing condition}
\label{app:simple}

Here I construct the boundary state $|P\rangle$ for
the boundary condition~\eqref{eq:simple_bc1},
using explicitly the gluing conditions~\eqref{eq:gluing_n}
and \eqref{eq:gluing_m}.
Although the boundary state can be given by
a more systematic geometric formulation as in Sec.
the present construction would be instructive
to understand the importance of the gluing conditions.

As I have discussed in Sec.~\ref{sec:simple},
the relevant boundary condition
would correspond to the Dirichlet boundary condition for $\Phi_1$,
namely $\Phi_1 = 0$, and the Neumann boundary condition
for $\Phi_0$.
The latter is equivalent to the Dirichlet boundary
condition for the dual field, $\Theta_0 = 0$.
Following the standard construction of the
Dirichlet/Neumann boundary state for
a free boson field theory,
I can write down an ansatz:
\begin{equation}
 | P \rangle = g_P \sum_{n_0,m_1} | (n_0, 0, 0, m_1) \rangle \rangle,
\end{equation}
where $|(n_0,n_1,m_0,m_1)\rangle$ is the Ishibashi
state with the winding numbers along the boundary
\begin{align}
 \Delta \Phi_j &= 2 \pi n_j \frac{R}{\sqrt{2}} ,\\
 \Delta \Theta_j &= 2 \pi m_j \frac{1}{\sqrt{2}gR},
\end{align}
where $j=0,1$.
The gluing conditions~\eqref{eq:gluing_n} and~\eqref{eq:gluing_m}
imply, because $n_1 = m_0 =0$, that $n_0$ and $m_1$ must be
even integers.
Thus the partition function (the amplitude with $|P\rangle$ boundary
state at the both ends) reads
\begin{equation}
 Z_{PP} = (g_P)^2 \left( \frac{1}{\eta{(\tilde{q})}}\right)^2
\sum_{n,m} \tilde{q}^{\frac{1}{2}( gR n^2+ \frac{m^2}{gR})} .
\end{equation}
Modular transforming, I obtain
\begin{equation}
 Z_{PP} = (g_P)^2 \left( \frac{1}{\eta{(q)}}\right)^2
\sum_{n,m} q^{\frac{1}{2}( gR n^2+ \frac{m^2}{gR})} ,
\end{equation}
which implies $g_P=1$ due to the Cardy's consistency condition.
Namely, there is no ``boundary entropy'' for this boundary,
as it is required by physical grounds.
Moreover, the partition function $Z_{PP}$ indeed agrees
exactly with the original expression~\eqref{eq:Zsimple}.

\bibliographystyle{iopart-num}
\bibliography{mybib}

\end{document}